# Spatial-Temporal Dynamics of High-Resolution Animal Social Networks: What Can We Learn from Domestic Animals?

Shi Chen, Amiyaal Ilany, Brad J. White, Michael W. Sanderson, and Cristina Lanzas

**Summary**

1. Recent studies of animal social networks have significantly increased our understanding of animal behavior, social interactions, and many important ecological and epidemiological processes. However, most of the studies are at low temporal and spatial resolution due to the difficulty in recording accurate contact information.

2. Domestic animals such as cattle have social behavior and serve as an excellent study system because their position can be explicitly and continuously tracked, allowing their social networks to be accurately constructed.

3. We used radio-frequency tags to accurately track cattle position and analyze high-resolution cattle social networks. We tested the hypothesis of temporal stationarity and spatial homogeneity in these high-resolution networks and demonstrated substantial spatial-temporal heterogeneity during different daily time periods (feeding and non-feeding) and in different areas of the pen (grain bunk, water trough, hay bunk, and other general pen area).

4. The social network structure is analyzed using global network characteristics (network density, exponential random graph model structure), subgroup clustering (modularity), triadic property (transitivity), and dyadic interactions (correlation coefficient from a quadratic assignment procedure). Cattle tend to have the strongest and most consistent contacts with others around the hay bunk during the feeding time. These results cannot be determined from data at lower spatial (aggregated at entire pen level) or temporal (aggregated at daily level) resolution.

5. These results reveal new insights for real-time animal social network structure dynamics, providing more accurate descriptions that allow more accurate modeling of multiple (both direct and indirect) disease transmission pathways.

**Introduction**

Analysis of animal social networks is essential to predict animal behavior, characterize animal social networks, and investigate many interesting ecological and evolutionary processes, including disease transmission (Croft et al. 2008). Many aspects of animal social networks change through time, from global network structure to basic dyadic (between two individuals) interactions (Blonder et al. 2012, Pinter-Wollman et al. 2014). There are two basic types of temporal dynamics on the network: the network structure (intrinsic changes in the network structure), and flow dynamics on the network (e.g. disease transmission, substance and energy flow in food webs)(Blonder et al. 2012). These two types of temporal dynamics, however, are not mutually exclusive as flow dynamics are also closely dependent on network structure changes, and vice versa. A static description of the network may not be able to accurately capture the comprehensive network structure and flows (Bansal et al. 2007, Fefferman & Ng 2007).

Recent studies have progressed in developing both concepts and techniques for temporal network dynamics. Temporal network dynamics can take place at different scales: at entire network/global level (how the global network characteristics, such as network density varies over time), at sub-graph level (how clusters of individuals form and change), at triadic level (how the transitivity in the triad changes among three vertices), and at dyadic level (how the relationship of each pair of individuals in the network evolves). Meanwhile, novel computational tools have also been developed to detect and quantify temporal network dynamics, such as quadratic assignment procedure (QAP, Krackhardt 1988), exponential random graph model (ERGM, Fienberg & Wassrman 1981, Robins et al. 2007, Hunter et al. 2008), modularity detection (Newman 2006, Fortunato 2010), transitivity test (Opsahl & Panzarasa 2009), and network comparison across space and time (Faust & Skvoretz 2002, Faust 2007).

Temporal networks have been applied to various wild animal populations, from primates to insects, including cyclic population structure in female African baboons (Henzi et al. 2009), social niche stability in primates (Flack et al. 2006), dynamics of social networks among Asian elephants (de Silva et al. 2011), social system restructure and dynamics in dolphins (Ansmann et al. 2012, Cantor et al. 2012), social network and genetic structure in spotted hyenas (Holekamp et al. 2012), social cohesion of marmots remaining at home (Blumstein et al. 2009), social network evolution and kin cluster in manakin birds (McDonald 2007, 2009), dynamic social networks in guppy fish (Croft et al. 2004, Wilson et al. 2014), temporal and individual variation in social ant colonies (Blonder & Dornhaus 2011, Pinter-Wollman et al. 2011).

A major problem for studies in wildlife ecology is that most of the wild animals are difficult to track individually and continuously with a high sampling rate (resolution) hence the resulting networks are at low temporal resolution, usually at daily level or longer. Furthermore, spatial heterogeneity is also important in animal social networks but often untraceable or ignored (summarized in Pinter-Wollman et al. 2014). These difficulties substantially hinder the utilization of temporal networks in ecology. These problems can be overcome in studies of networks for domestic animals. Cattle, as for many other ungulates, present social interactions such as exploratory behavior, recognition, communication (especially tactile communication), and peer bonding (Philips 2002). Although quantitative descriptions of cattle social networks are still uncommon, dynamic social networks have been recorded in both dairy and beef cattle populations (Gygax et al. 2010, Chen et al. 2013, 2014).

Cattle are usually in a confined area so that their position can be explicitly and accurately tracked in order to construct ultra-high resolution social networks (at second and centimeter resolution temporally and spatially), thus the completed network can be accurately described. Spatial heterogeneity also plays an important role in network dynamics (Peletier et al. 2003, Peterson et al. 2013), and data at high-resolution for cattle social networks can facilitate better understanding of the complicated spatial-temporal dynamics (Chen et al. 2014). Furthermore, the group size of a cattle population is generally consistent (excluding immigration and emigration, as opposed to wild animals), thus allowing for some computational analysis (e.g. QAP which usually requires the same set of individuals/vertices in the networks). Thus cattle are suitable candidate to study spatial-temporal dynamics of animal social networks that can provide insight on the dynamics of such networks in other domestic and non-domestic species.

We here construct high-resolution cattle social networks accounting for both temporal (at hourly level in a day) and spatial (different areas in the pen, e.g. feeding bunk, water supply, and hay bunk) heterogeneities. We comprehensively and quantitatively compare and contrast how network structure changes at different levels, from global network characteristics, to subgraph, triadic, and down to the dyadic interactions, at different time and area, and demonstrate the spatial-temporal heterogeneity at these different levels in an cattle social network. This enhances our understanding of the real-time animal social network dynamics, develops more realistic description and characterization of social networks, and facilitates future research such as incorporating accurate disease transmission models on real-time social networks.

**Materials and Methods**

Data Acquisition and Standardization

Real-time animal position data were collected at the research farm of Kansas State University (Manhattan, KS, USA). Three pens of Holstein calves were monitored, with 21, 21, and 27 calves in pens #1 to #3, respectively. The entire pen is partitioned into four exclusive areas: around the grain bunk ($5m^2$), the hay bunk ($5m^2$), and the water trough ($3m^2$), and all other general pen floor ($187m^2$). Each calf was continuously monitored from August 11, 2011, to August 18, 2011, using an wide-band radio frequency tag (Ubisense Series 7000 Compact Tag, Ubisense) attached to its ear. The tag transmitted ultra wide-band signals to seven receivers around the pens; the receivers then transferred data to a central computer that logs the 2-dimensional (X-Y coordinates) position data. This monitoring system was accurate (up to 0.01 m in spatial resolution) and did not disturb the animals nor change their normal behavior. The original position data reported by the tags of each animal were aggregated to 10-s resolution. Hence, in each day, each animal would have 8,640 2-dimensional data points recording its location (24 h/d×60 min/h×6 data/min). The detailed data acquisition and standardization processes are presented in our previous work (Chen et al. 2013, 2014) as well as in the supplementary information (SI) with this manuscript.

Network Construction

First, the complete cattle-cattle social network (which does not take spatial areas into consideration) is generated based on the cattle position data in each hour using a 0.3m threshold distance, as described in Chen et al. (2014). Then whenever a direct contact occurs, the positions of the pair are determined and such direct contact is attributed to one of the four areas, using the same 0.3m threshold distance. For example, if two cattle have a contact and their position falls within the 0.3m threshold distance of the grain bunk, this direct contact occurs around the grain bunk. Occasionally, a direct contact can be attributed to more than one area: consider a pair of cattle in contact with animal $i$ within threshold distance to the grain bunk but animal $j$ at the general pen floor. In this situation, priority is given to identify the contact as being associated with the more specific areas (e.g. grain, hay, and water) than the general pen floor, and this contact is assumed to occur at the more specific area. Note that grain, hay, and water themselves in the study pen are sufficiently spatially separated that no direct contacts can occur simultaneously in two of them.

The complete social network is partitioned to four spatially-explicit networks representing four different areas in the pen in each hour. These social networks are weighted networks (i.e. the values in the adjacent matrix $M$ are not binary but assigned numerical values for the frequency of contacts). For example, if the value of $M_{ij}$ is $m_{ij}$, it indicates that two cattle $i$ and $j$ have a total of $m_{ij}$ contacts within that hour. In this study we assume an undirected network such that $M_{ij} = M_{ji}$ (animal $i$ contacting animal $j$ implies animal $j$ contacting animal $i$ ), so $M$ is a symmetric matrix. The indirect contact structure (i.e. number of contacts between an individual animal and specific areas such as the grain, hay, and water) is also obtained on an hourly basis using the same 0.3m threshold distance, and the time series of indirect contact structure is calculated for these specific areas.

As shown later in the results section, the regular feeding activity (twice a day around 8-9am and 2-3pm, shortened as 8am and 2pm hereafter, and food is supplied at the grain bunk) significantly increases cattle contacts during these periods. Our observations also show cattle usually gather around hay to ruminate shortly after feeding. Thus in each day the feeding time (8am and 2pm) is separated from all other non-feeding time. We first investigate the potential coupling and synchronization between indirect contact (with grain, water, and hay) and direct contact using time series analysis and spectrum analysis. This step is necessary to reveal temporal heterogeneity in network structure and its potential cause. Then we explicitly and comprehensively characterize and compare the network structure dynamics at different time of day and different areas in the pen to understand the spatial-temporal heterogeneities in the contact net.

Network Characterization and Comparison

Many properties and statistics are used to characterize the networks from a global level down to dyadic interactions. The network density is one of the most important and common measure of networks at a global level. In our study system (weighted network), the network density is proportional to the total number of contacts in each hour. The network density is bracketed between 0 and 1, where 0 indicates no contact in that time period (1-h) at all and 1 is associated with a fully connected network (i.e. each pair of vertices is connected). However density alone is insufficient to simulate networks with similar structure to observed networks. To deliver more robust statistical inference on the network, we also fit the observed networks with ERGMs to the edges, vertex covariate (e.g. gender of cattle), and other network structures (e.g. triangle, k-stars, isolates, see Hunter et al. 2008), and obtain the estimates of these parameters. The null model only consists edges, and the final model is chosen based on the smallest AIC (Akaike Information Criterion) value and goodness-of-fit test. Particularly, the edge coefficient of ERGMs ($\theta$) is an important characteristic of the network known as conditional odds. The probability of forming an edge is given by $p=e^{\theta}/(1+e^{\theta})$. Details of procedures in ERGM are presented in Robins et al. (2007) and Hunter et al. (2008).

Furthermore, the vertices in the network (in our study system, the cattle) can show a clustering pattern and form groups. Modularity measures the strength of the division into cluster/groups. Modularity can be positive or negative, and a larger positive value indicates stronger clustering pattern (Newman 2009). At the triadic level (three vertices), transitivity, also known as the global clustering coefficient, measures which vertices tend to cluster together (Opsahl & Panzarasa 2009). The means and standard errors of modularity and transitivity are also obtained in these networks.

In order to compare network change over time and across space at the dyadic level (two vertices, i.e. whether two cattle consistently act together), the QAP is applied. The QAP correlation measures how dyadic interaction changes through time and a larger QAP correlation indicates more consistent and stable dyadic interaction between two networks at different times. As indicated before, the time (hour) of day is considered either "feeding" or "non-feeding". For all the 192 observed networks for each area, there are a total of 16 networks during feeding time (8am/2pm, for 8 days) and 176 for non-feeding time. The QAP is applied between each pair of networks during feeding time (a total of 16*15/2=120 pairs), between each pair of networks during non-feeding time (a total of 15,400 pairs), and between each pair of networks in feeding and non-feeding times (a total of 1408 pairs). All these QAP correlation values are compared between different time periods of day as well as for different areas in order to reveal dyadic interaction changes. All the analyses are performed in *R* 3.1.0, with the *statnet* meta-package.

**Results**

The time series of cattle social network density (proportional to number of contacts) for the complete observation period (192h in total) is plotted against the time series of indirect contact between cattle and grain, water, and hay, respectively (Fig. 1). The time series of direct contact show substantial diurnal cycle, and the time series between direct and indirect contacts (grain and hay) are significantly coupled, which is supported by the coherency plot (since the solid black coherency line is usually above the red significance level). However the time series between direct and indirect contact with water are not significantly correlated (the coherency line hardly exceeds the significance level). These results show that cattle social network density changes significantly within a day, and feeding activity promotes clustering (thus more dense social networks) around grain and hay, while drinking is not a key factor for network changes. These results indicate that there is neither temporal stationarity nor spatial homogeneity in this high-resolution cattle social network (number of contacts).

Next, we show four important network measurements (density, ERGM coefficient, modularity, and transitivity, covering levels from the global network structure to dyadic interactions) in different time periods (feeding and non-feeding) and in different areas (grain, water, hay, and other pen floor) to explicitly demonstrate the spatial-temporal heterogeneity. The network density measures at a global network structure level and can be approximated as the number of total contacts. It is not surprising that network density is much higher during feeding time than non-feeding time (Fig. 2, $1^{st}$ row), as supported by the time series plot in Fig. 1 as well. Interestingly, the cattle have more contacts around hay than grain even during feeding time. This can be explained by the fact that the food supply at the grain is limited and feeding behavior usually only lasts for 15-20 minutes. However the cattle spend significantly longer time (about 45-60 minutes) at hay after feeding to re-ruminate, and during rumination the cattle tend to gather together. These findings of spatial and temporal global network structure changes are also supported by the more robust inference from the ERGM. The best-fit ERGMs for all the four areas during feeding time have effects from edge, gender, and triangle (with all parameters converged in the ERGM procedure); while non-feeding time have edge, gender, and isolates (not all parameters converged possibly due to the 0-inflated matrix structure). Specifically for edge effect, the function describing edge formation ($e^{\theta}/(1+e^{\theta})$) monotonically increases with $\theta$ and positive edge coefficients correspond to larger probability (>0.5) of forming edges in the network. Networks during feeding time around hay and other pen floor have positive ERGM coefficients while negative ones around grain and water (Fig. 2, $2^{nd}$ row, left panel). As a comparison, all the ERGM coefficients are negative and do not differ much during the non-feeding time (Fig. 2, $2^{nd}$ row, right panel). The values of ERGM coefficients also coincide with network density results such that edges are more frequently formed during feeding time around hay. For the grouping and clustering of the networks, results reveal very low modularity value (almost 0) around hay during feeding time, which indicates almost all cattle have some connection with others (Fig. 2, $3^{rd}$ row, left panel). This is very distinct from around the grain, water, and hay during feeding time, and from all areas during non-feeding time.

During feeding time, water area has the largest modularity value (Fig. 2, $3^{rd}$ row). Although the modularity values indicate cattle have some connection with others at hay during feeding time, it is then revealed by the transitivity values that they actually divide into more stable subgroups at hay during feeding time (Fig. 2, $4^{th}$ row, left panel). This paradoxical result can be explained by the fact that during feeding time, the cattle tend to cluster into several subgroups and these subgroups are not isolated around hay (i.e. different subgroups are connected by some "bridge" vertex/individual). Such a network shows low modularity (i.e. vertices are more or less connected) but high transitivity (i.e. the vertices form distinctive subgroups). The transitivity values are not significantly different among different areas during non-feeding time. Sample illustrations are given in Fig. 3 showing the real social networks and clusters in four different areas during both feeding and non-feeding time.

At dyadic level, the correlations are much stronger (larger QAP correlation values) between any two feeding time periods (FF) than between any two non-feeding time periods (NN) and between one feeding and one non-feeding time periods (FN), which indicates more stable and consistent dyadic interactions during feeding time at all four locations (Fig. 4). Thus feeding tends to promote and preserve the network structure. For different areas during feeding time, hay has significantly higher QAP correlation value (0.10) than grain (0.03), water (0.02), and all other pen floors (0.02), meaning cattle tend to interact with others more consistently around hay. All these results have demonstrated significant temporal and spatial heterogeneity in the cattle social networks.

**Discussion**

Our results clearly reveal that the cattle social network has substantial spatial and temporal heterogeneity. The network structure, from global characteristics to basic dyadic interactions, changes during different times of the day as well as in different areas. Most of the contacts (approximately 60% of total contacts) happen during the feeding time (a total of 2 hours per day, which accounts for only 8% of total daily time). Interestingly, hay accounts for more direct contacts during feeding time than grain, as can be seen from the network density plot (Fig. 2). It also coincides with our previous finding that cattle spend significantly longer time around hay (about 2 hours per day) than around the grain bunk (about 1 hour per day, Chen et al. 2013). These results show the possible correlation between direct animal-animal contacts and indirect animal-environment contacts. However, the spatial-temporal heterogeneities in the cattle social network diminish when the networks are aggregated at lower temporal (e.g. at daily level or above) and spatial (e.g. aggregated at entire pen) resolution. A major challenge to current studies of dynamic networks is lack of sufficiently resolved data, especially for wildlife studies (Blonder et al. 2012, Holme & Saramaki 2013a, 2013b, Krause et al. 2013, Kurvers et al. 2014). Thus cattle (and other domestic animals) serve as an excellent example to study and understand real-time evolution of network structure and potentially shed light on wildlife studies, especially the optimal observation strategy. For instance, researchers can observe animal social network during specific short period of time while preserving most of the network structure characteristics. Besides, the results from ERGM reveal that adding information of individual animal characteristic (in this study, gender information) leads to better fit and characterization of network structure (i.e. the ERGM model with gender information has smaller AIC value than the model without such information).

Except very specific behavior interactions, many animal social networks are constructed based on the assumption that spatial proximity reflects social closeness. But this assumption may be incorrect because distance alone (and hence contact) does not necessarily reflect real social ties (Pinter-Wollman et al. 2014). Adult cattle have approximately 2m flight distance (an animal will change normal behavior and move if get closer than the flight distance) and young calves have smaller flight distance (Philips 2002). When a young calf is less than the threshold distance (0.3m in this study) from other calves, it does not necessarily imply the two calves are having some social interaction – they may be just randomly passing by each other. The results, especially QAP values, from our study show that the dyadic interaction is unstable and inconsistent during non-feeding time periods. Thus, we argue that the some of the contacts during non-feeding time are not real social interactions but perhaps just random contacts. During feeding time, social networks formed around hay are much more stable than at the grain bunk. We suggest that during feeding time, the cattle need to compete with others for food in the grain bunk thus cannot always eat with an intentionally chosen partner. On the other hand, after feeding the cattle can go with their partner freely to the hay for feeding. Thus the contacts around the grain bunk are facilitated by feeding activity but may not necessarily reflect social ties. It is only the contacts around the hay bunk during feeding time that may attribute to the real social ties. Thus based on the spatial and temporal heterogeneity in the social network, we can categorize the contacts into three types: social, random, and indirect-contact facilitated contact. Our study provides a potential way to infer the real social network based on the observed social network.

One of the major applications of dynamic networks is to model disease transmission on the social networks. Recent studies have already revealed the importance of spatial-temporal heterogeneity in the contact structure for disease transmission in direct animal-animal contact pathway and indirect animal-environment pathway (Chen et al. 2013, 2014). Synthetic data are used to simulate multiple transmission pathways simultaneously (Cortez and Weitz 2013) but the different pathways are considered uncorrelated as well. Our results demonstrate that indirect contact between animal and environment and direct contact between animals can be highly synchronized and coupled (especially during feeding time). This enables future research on modeling multiple transmission pathways at the same time, explicitly investigating the relative importance of each transmission pathway (e.g. direct contact, foodborne, waterborne, etc.), and measuring the effects of spatial-temporal heterogeneity with a realistic high-resolution social network.

**Acknowledgement**

We thank the comments and suggestions from Dr. Louis Gross, National Institute for Biological and Mathematical Synthesis (NIMBioS), for improvements of this manuscript. This work was conducted with partial funding provided at NIMBioS, an institute sponsored by the U.S. National Science Foundation, the U.S. Department of Homeland Security, and the U.S. Department of Agriculture through NSF Award # EF-0832858, with additional support from the University of Tennessee, Knoxville. The authors declare no conflicts of interests in this study.

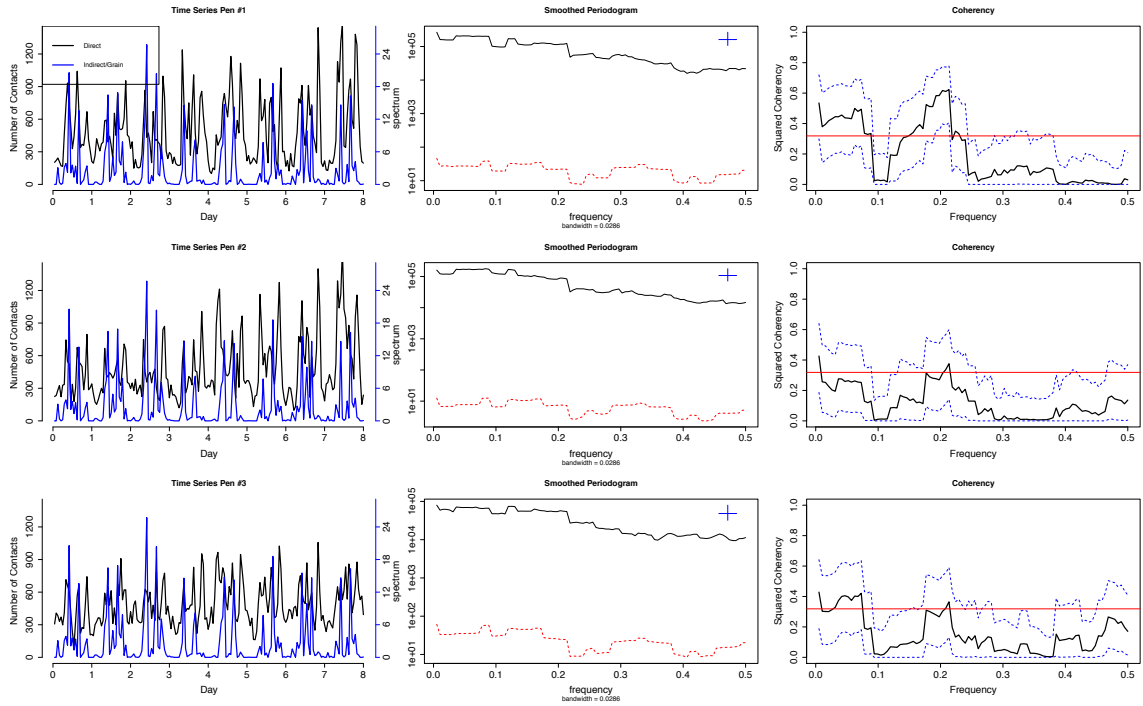

**Fig. 1A**

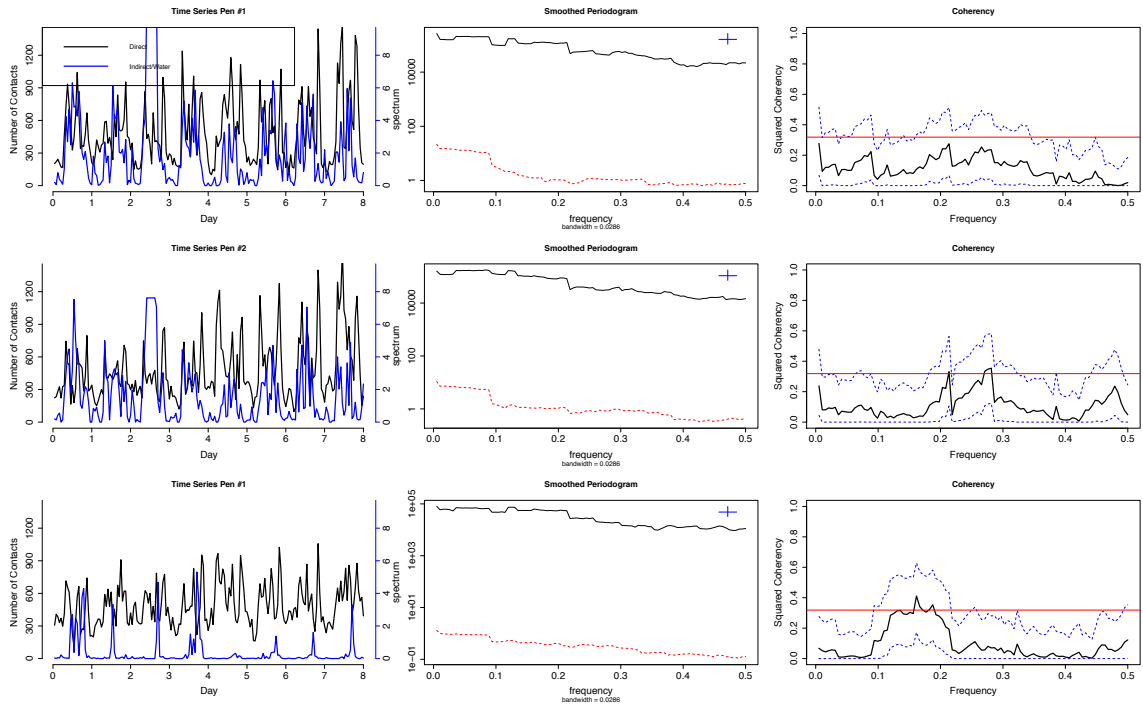

**Fig. 1B**

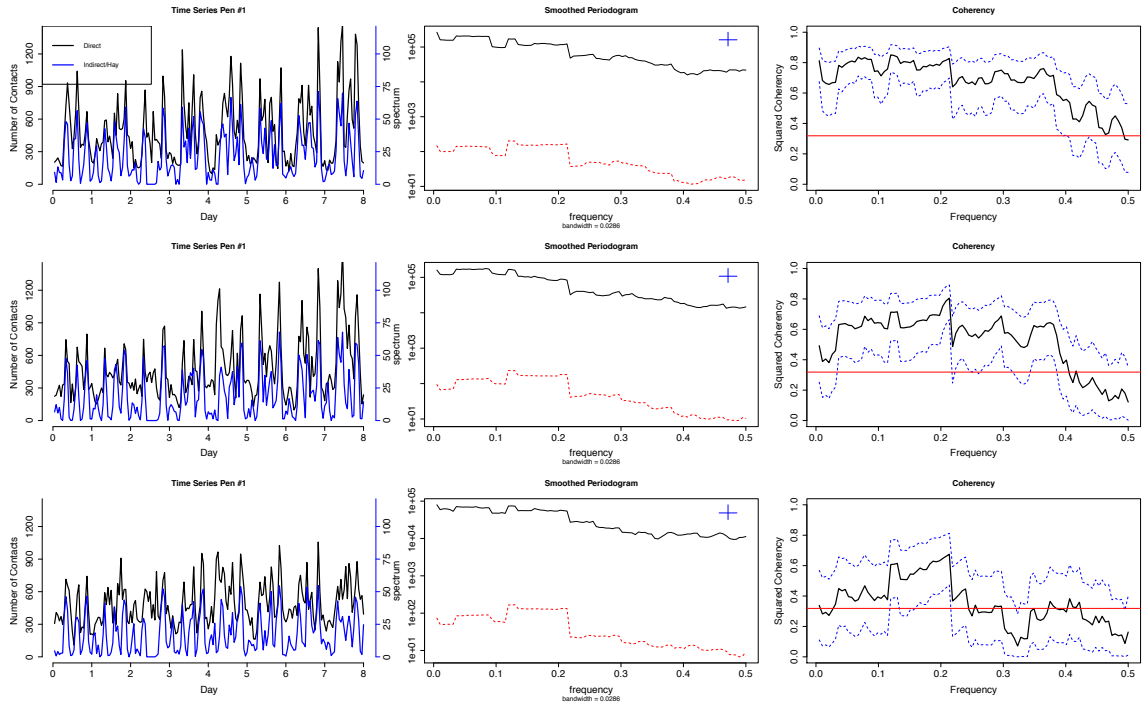

**Fig. 1C**

**Fig 1. Time Series of Indirect Contact (Grain, Water, Hay), Periodogram, and Coherency Plot**

Note: Fig. 1A, 1B, 1C are for grain, water, and hay, respectively. The time series of indirect contact and direct contact show high synchrony (coupling) in grain and hay, as the coherency plots (solid black line) have substantial part above the threshold (solid red line) while there is no clear relationship between indirect contact with water and direct contact.

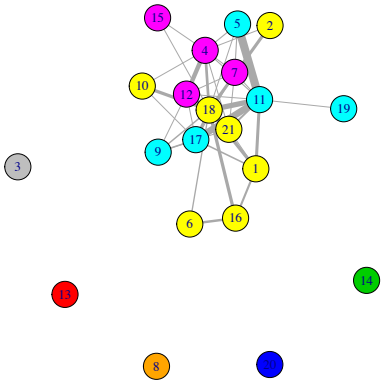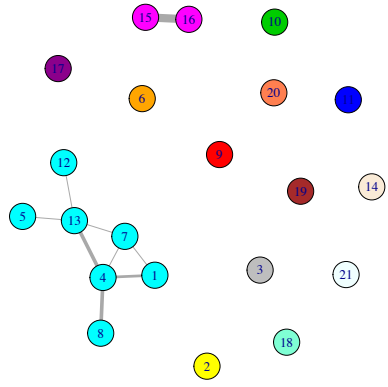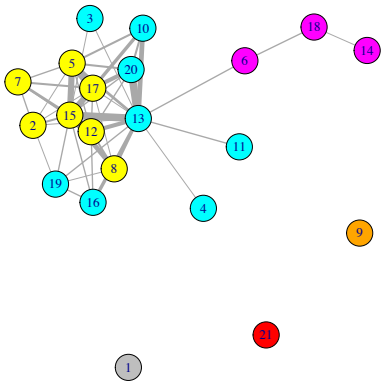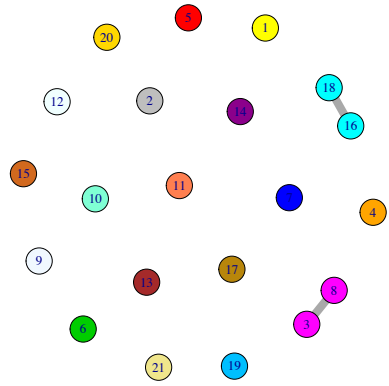

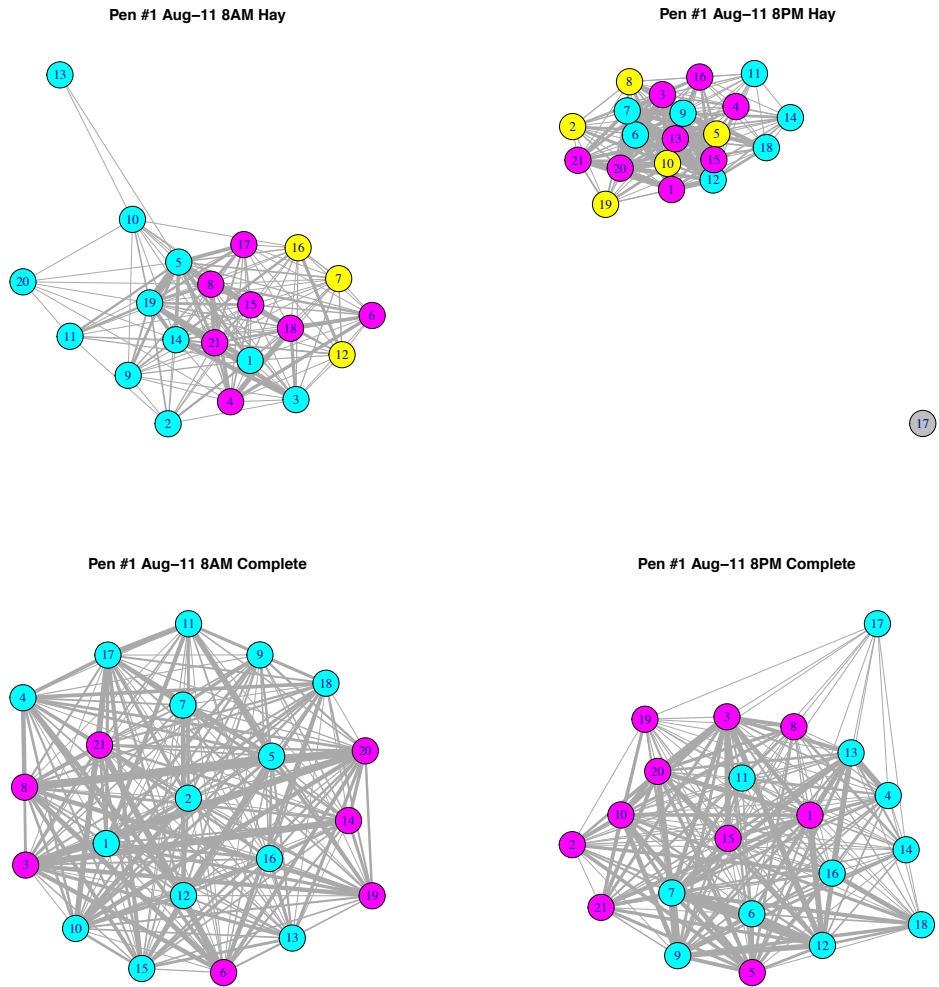

**Fig 2. Network Structure with Modularity Examples at Different Time and Different Area**

Note: from top to bottom: networks around grain, hay, water, and complete network in the pen. Left: networks aggregated during 8-9AM on Aug. 11. Right: networks aggregated during 8-9PM on Aug. 11. Different colors indicate modular (subgraph) based on the modularity analysis in *R*. Thickest line in each network corresponds to the largest number of contacts for that particular area in one hour, thus not directly comparable between different networks. Networks in grain, hay, and water show strong spatial-temporal heterogeneity and distinctive modularity, but such heterogeneity diminish when aggregated to the complete networks.

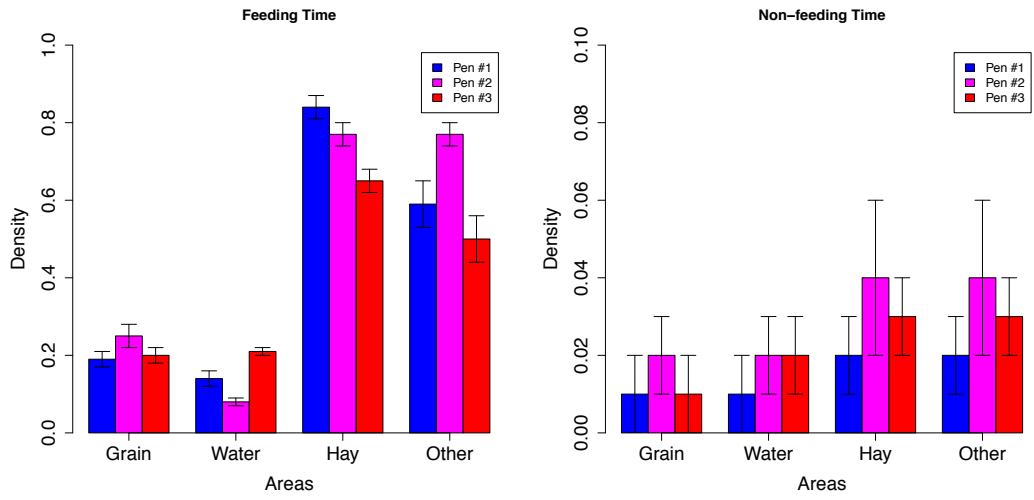

**Fig. 3A**

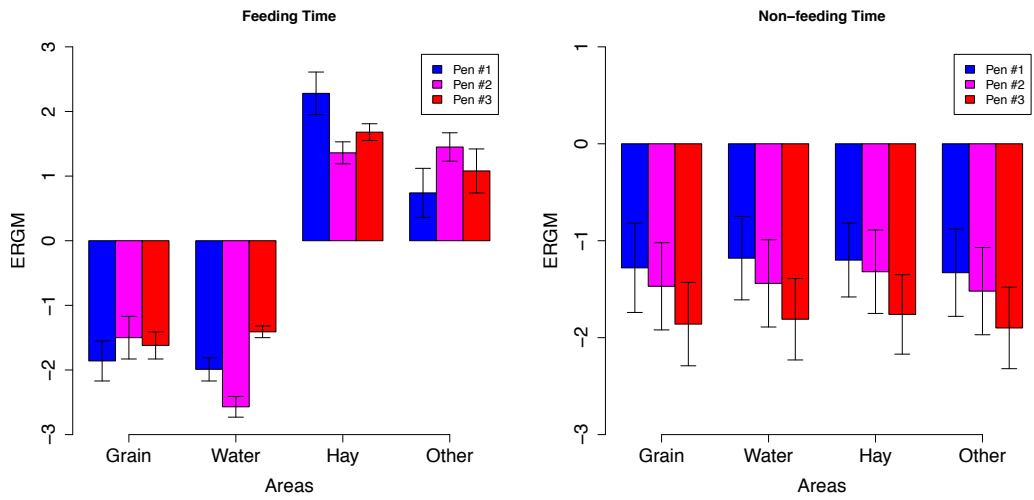

**Fig. 3B**

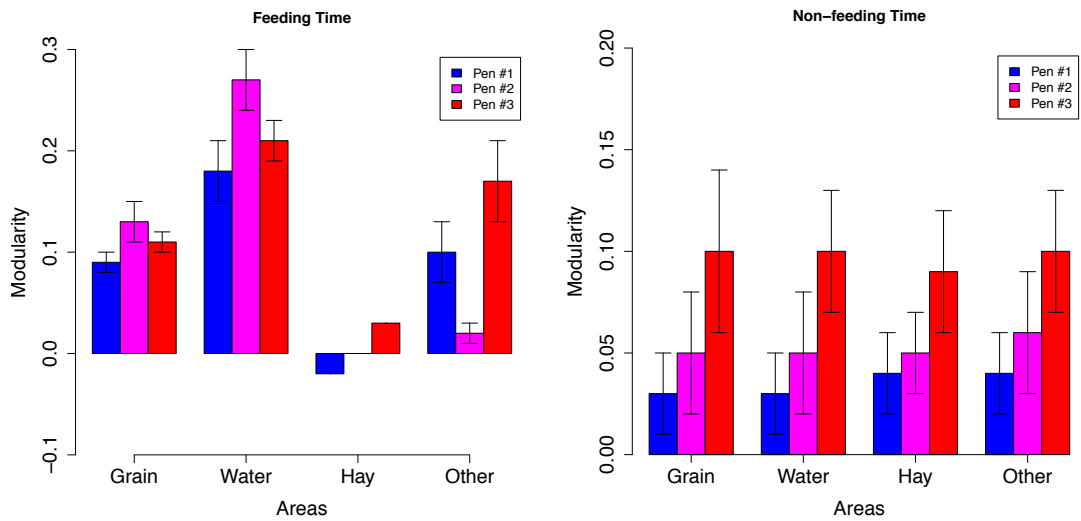

**Fig. 3C**

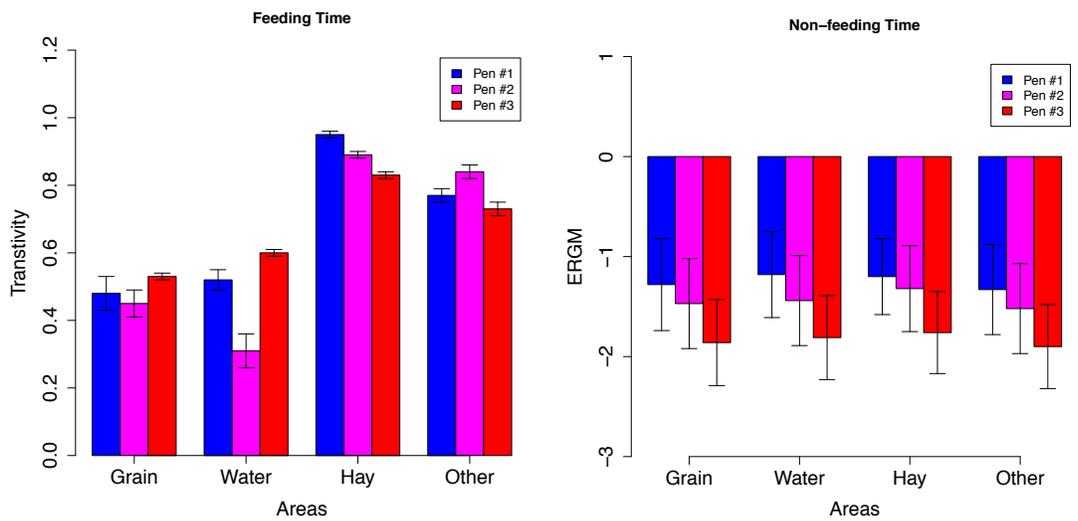

**Fig. 3D**

**Fig 3. Network Structure Characteristics (density, ERGM coefficient, modularity, and transitivity) at Different Time and Different Area**

Note: Fig. 3A, 3B, 3C, 3D are for density, ERGM coefficient, modularity and transitivity values, respectively. Left: feeding time (between 8AM-9AM and 2PM-3PM); right: non-feeding time (other time periods). Note all the network characteristics differ significantly between feeding and non-feeding time (left vs right panel in each sub-figure). Spatial heterogeneity (different areas) does not have significant impact during non-feeding time but substantially alter network characteristics during feeding time.

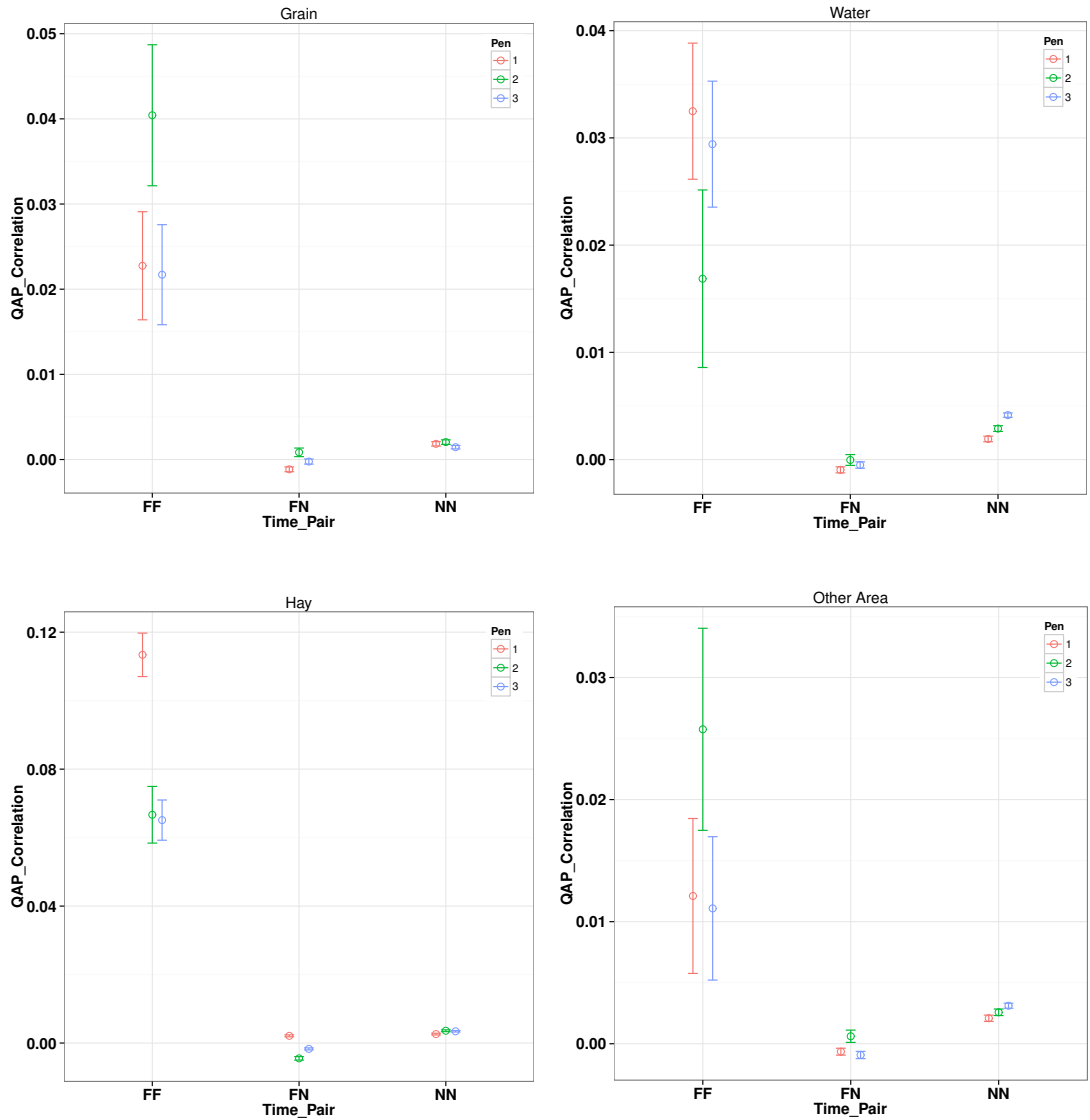

**Fig 4. QAP Correlation Test Results for Different Time Period Pairs at Different Areas**

Note: the four sub-figures are for grain, water, hay, and all the other areas, respectively. F indicates feeding time period (8-9AM, 2-3PM) and N indicates non-feeding time periods. Within each subfigure (the same area) QAP is significantly larger between two feeding time periods (FF) than between one feeding and one non-feeding time periods (FN) and between two non-feeding time periods (NN). Hay has the much larger QAP correlation during feeding time (FF, valued at about 0.1) than grain (0.03), water (0.02), and all the other area (0.02). Thus the contact network structure especially dyadic interaction is most consistent and stable during feeding time around hay.